\documentstyle[prl,aps,psfig,multicol]{revtex}
\begin{document}
\draft
\title{Response to Parallel Magnetic Field of a
Dilute 2D Electron System\\ across the Metal-Insulator Transition}
\author{ K.~M.~Mertes, D.~Simonian\cite{ds}, and M.~P.~Sarachik}
\address{Physics Department, City College of the City
University of New York, New York, New York 10031}
\author{S.~V.~Kravchenko}
\address{Physics Department, Northeastern University,
Boston, Massachusetts 02115}
\author{T.~M.~Klapwijk}
\address{Delft University of Technology, Department of Applied Physics,
2628 CJ Delft, The Netherlands}
\date{\today}
\maketitle
\begin{abstract}
The response to a parallel magnetic field of the very dilute insulating
two-dimensional system of electrons in silicon MOSFET's is dramatic and
similar to that found on the conducting side of the metal-insulator
transition:  there is a large initial
increase in resistivity with increasing field, followed by saturation to
a value that is approximately constant above a characteristic magnetic
field of about one Tesla.  This is unexpected behavior in an insulator that
exhibits Efros-Shklovskii variable-range hopping in zero field, and appears
to be a general feature of very dilute electron systems.
\end{abstract}
\makeatletter
\global\@specialpagefalse
\def\@oddhead{REV\TeX{} 3.0\hfill To be published in {\it Phys.\ Rev.\ B}}
\let\@evenhead\@oddhead
\makeatother
\begin{multicols}{2}

Until quite recently, it was believed that all two-dimensional systems
of electrons (or holes) are necessarily localized in the absence
of a magnetic field in the limit of zero temperature.  This conclusion was
based on the scaling theory for non-interacting electrons of Abrahams
{\it et al.} \cite{abrahams79}, was
further confirmed theoretically for weakly interacting electrons
\cite{altshuler80,caveat}, and received experimental confirmation in a number
of materials, including thin films \cite{osheroff} and (high-density) silicon
metal-oxide-semiconductor field-effect transistors
(MOSFET's)\cite{pepper,bishop}.
In the last several years, however, measurements in very dilute
two-dimensional systems have provided evidence
of a transition from insulating to conducting behavior with
increasing electron (hole) density above some low critical value on
the order of $10^9$ to $10^{11}$ cm$^{-2}$
\cite{kravchenko,popovic97,coleridge97,hanein98,simmons98,papadakis98,hanein98a}
.  At these very low densities the energy of
electron-electron interactions exceeds the Fermi energy by an order of
magnitude or more, and correlations thus
provide the dominant energy in the problem.  Dilute, strongly
interacting
two-dimensional systems are currently the focus of intense theoretical
interest, and have elicited a spate of theoretical attempts to account
for the presence and nature of the unexpected conducting phase.

One of the most interesting characteristics of the conducting phase
is its dramatic response to a magnetic field applied parallel to the plane of
the two-dimensional system.  For example, the resistivity of very
high-mobility silicon MOSFET's increases by almost three orders of magnitude
with increasing field, saturating to a new value in fields above
$\sim2-3$~Tesla \cite{simonian97a,pudalovH}.  A similar effect
was observed in p-GaAs/AlGaAs heterostructures \cite{simmons98} confirming
that this giant positive magnetoresistance is a general property of dilute
conducting 2D systems\cite{wheeler}.  In Ref.~\cite{dolgopolov92}, it
was reported that the metal-insulator transition in Si MOSFET's shifts toward
higher electron densities in a parallel magnetic field of the order of a few
Tesla, while at higher magnetic fields, the effect saturates.  We note that
a parallel magnetic field
couples only to the spins of the electrons and not to their orbital motion.
Spins are thus known to play a crucial role, and it has been suggested that
full alignment of the electrons results in the complete suppression of the
anomalous conducting phase.

In this paper we report that the response of the very dilute 2D system of
electrons in high-mobility silicon MOSFET's to a parallel magnetic field is
qualitatively the same in the insulating phase, varying continuously for
electron densities spanning the transition from insulating to conducting
behavior.  This implies that spins play as crucial a role in the insulating
phase as they do in the conducting phase.

The silicon MOSFET's used in these studies were samples with split gates
especially designed for measurements at low electron densities and low
temperatures similar to those used previously in
Ref.~\cite{heemskerk98}.  The split gates allowed independent control
of the electron density in the main channel and in the contact
region, allowing a high ($\sim10^{12}$~cm$^{-2}$) electron
density to be maintained near the contacts to minimize contact resistance.
Sample mobilities at $T=4.2$~K were close to 25,000~cm$^2$/Vs.

The resistivity is shown in Fig.~1 on a logarithmic scale as a function of
temperature for different electron densities $n_s$ (determined by the voltage
applied between the gate and the 2D layer) spanning the metal-insulator
transition.  The data were taken using low-frequency (typically 0.5~Hz)
and low-current ac techniques at higher
\vbox{
\vspace{.8in}
\hbox{
\hspace{.15in}
\psfig{file=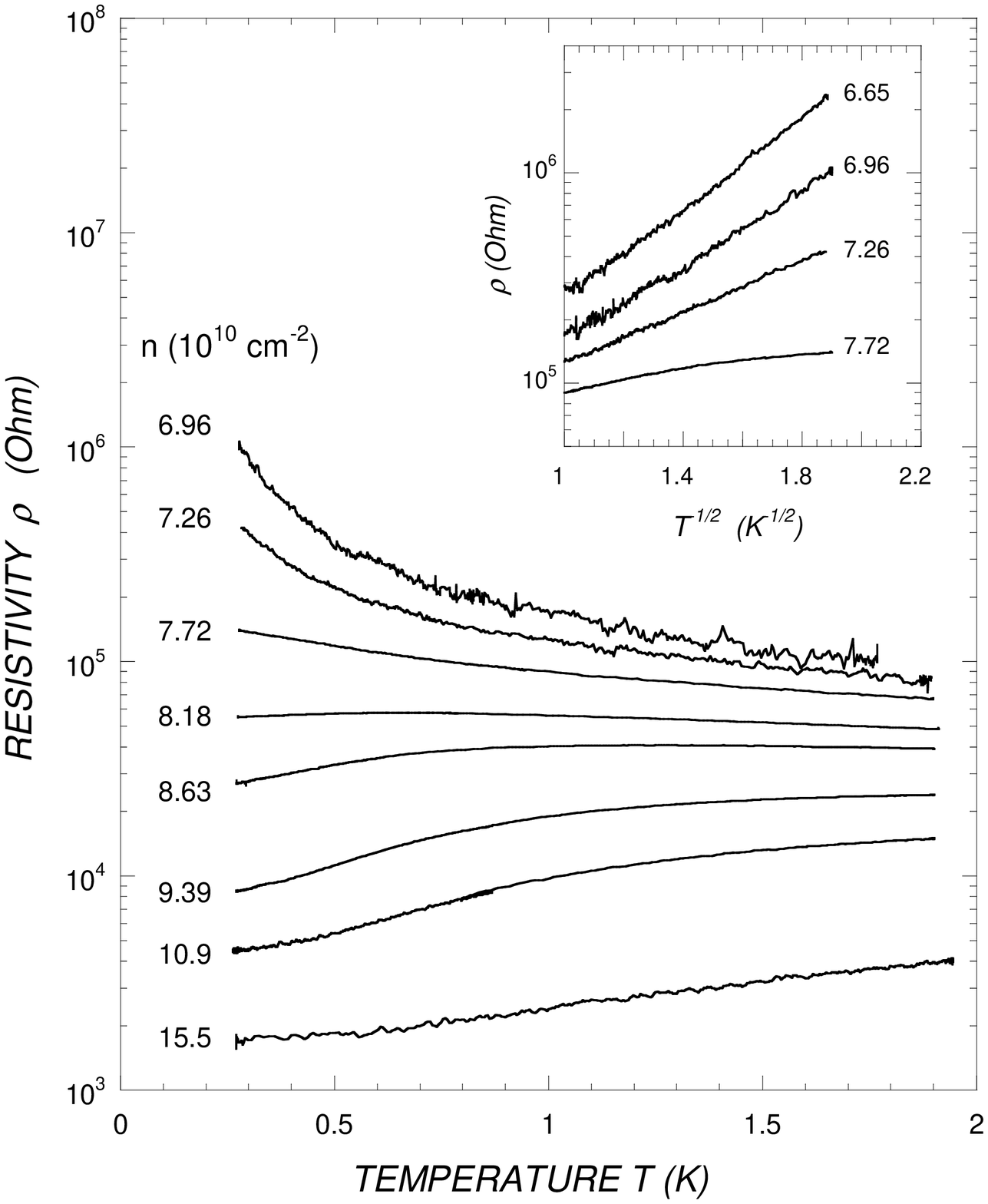,width=3.1in,bbllx=1.5in,bblly=1in,bburx=7.75in,bbury=8.5in
,angle=0}
}
\vspace{-0.15in}
\hbox{
\hspace{-0.15in}
\refstepcounter{figure}
\parbox[b]{3.4in}{\baselineskip=12pt \egtrm FIG.~\thefigure.
Resistivity in zero field as a function of temperature for
several electron densities, as labeled.  The critical density for the
conductor-insulator transition is $7.87\times 10^{10}$~cm$^{-2}$.  For
electron densities $n_s<n_c$, the inset shows the log of the resistivity
versus $T^{-1/2}$, demonstrating that Efros-Shklovskii variable-range
hopping is obeyed for low densities.
\vspace{0.10in}
}
\label{1}
}
}
densities (six lower curves) and
low-current dc techniques at lower $n_s$ (two upper curves); the latter
resulted in noisier data. In all cases, care was taken to ensure
linearity.  The resistivity increases (decreases) with decreasing
temperature
for low (high) electron densities, signaling a transition from insulating to
conducting behavior at a critical electron density $n_c=(7.87\pm0.10)\times
10^{10}\mbox{ cm}^{-2}$.  The upper two curves in the main figure are clearly
in the insulating phase and the third curve from the top is barely insulating.

For electron densities spanning the transition, Fig.~2 shows the
resistivity at
300~mK as a function of magnetic field applied parallel to the plane of the
electrons;  the top three curves are insulating in zero field while the
remaining curves are in the conducting phase.  Again,
low-frequency/low-current ac technique was used at the higher densities
(three lower curves) and a low-current dc technique was used at lower
$n_s$.  Measurements have indicated that the overall size
of the magnetoresistance is larger at low temperatures and for samples of
higher mobility.  As reported earlier \cite{simonian97a,pudalovH}, the
resistivity remains approximately constant at small fields, then rises
steeply with increasing field by more than an order of magnitude
\vbox{
\vspace{-0.07in}
\hbox{
\hspace{.6in}
\psfig{file=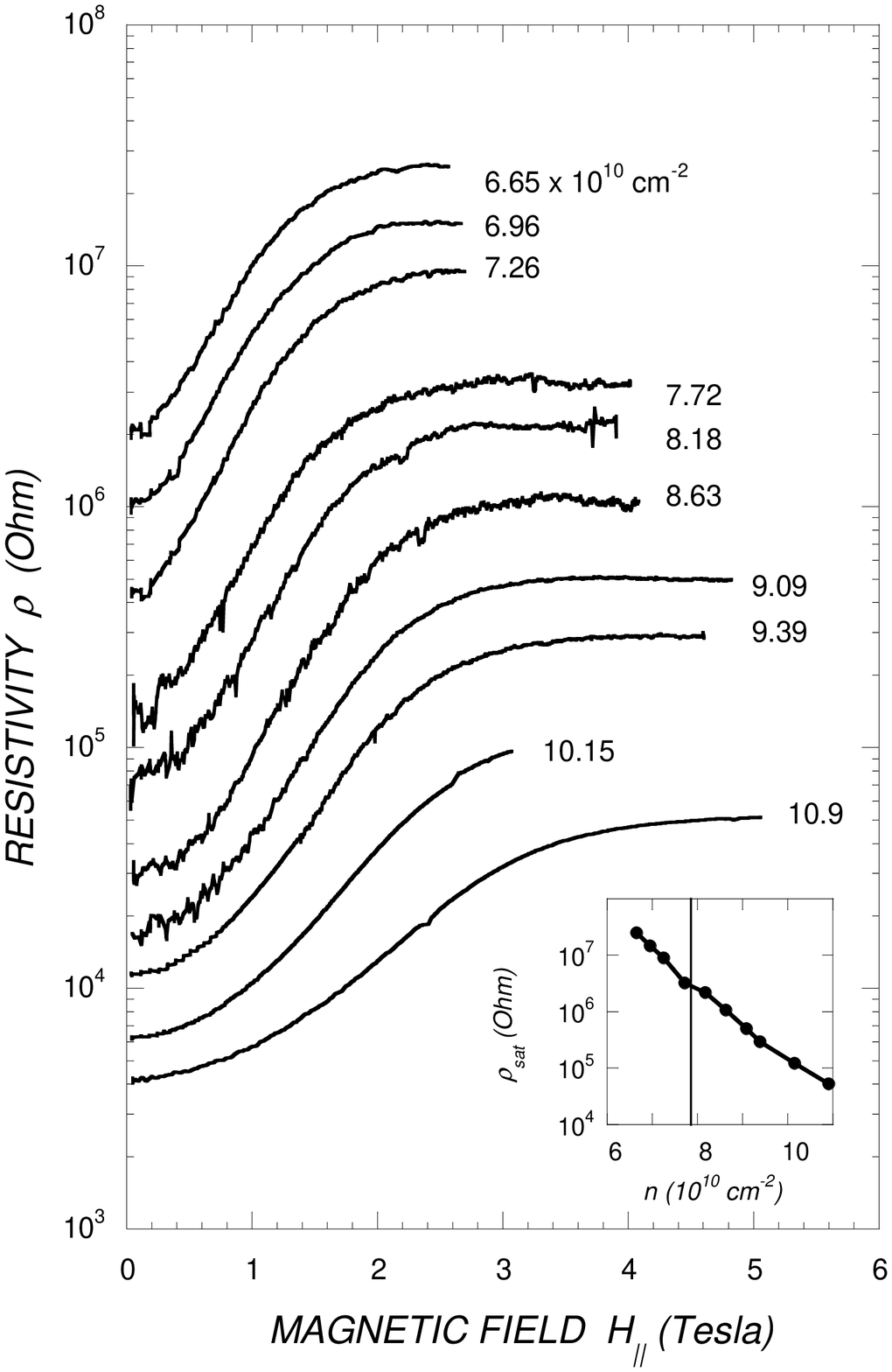,width=3.1in,bbllx=1.5in,bblly=1in,bburx=7.75in,bbury=10.5in
,angle=0}
}
\hbox{
\hspace{-0.15in}
\refstepcounter{figure}
\parbox[b]{3.4in}{\baselineskip=12pt \egtrm FIG.~\thefigure.
Resistivity at 300 mK as a function of in-plane magnetic field
for different electron densities spanning the conductor-insulator
transition.  The inset shows the resistivity at saturation, $\rho_{sat}$,
versus electron density; the vertical line denotes the critical density.
\vspace{0.10in}
}
\label{2}
}
}
(depending on
the electron density) and saturates to a new value for fields above about 2
or 3~Tesla, depending again on electron density.  The inset shows
$\rho_{sat}(H_{||})$ as a function of $n_s$; the vertical line denotes our
estimate for the critical density $n_c$.  The surprise here is that the
behavior of the magnetoresistance is essentially the same in the insulating
phase as in the conducting phase.  The response of the resistivity to parallel
field evolves continuously and smoothly, with no indication that a transition
has been crossed.  A similar giant increase of the resistivity
in response to an in-plane magnetic field has been observed recently by
Khondaker {\it et al.}\cite{shlimak} in insulating $\delta$-doped GaAs/AlGaAs
heterostructures.

In agreement with earlier reports \cite{mason95}, the inset to Fig.~1 shows
that in the absence of a magnetic field, the resistivity of these high-mobility
silicon MOSFET's for $n_s<n_c$ obeys variable-range hopping
of the Efros-Shklovskii (ES) form\cite{ES}, $\rho(T) =\rho_0\,$exp$
(T_0/T)^{1/2}$, where $\rho_0$ was found to be independent of temperature and
close to $h/e^2$.  As expected, departures from this form are evident at
higher temperatures (for $T^{-1/2} < 1.1$) as well as for the density
$n_s=7.72\times 10^{10}$~cm$^{-2}$ very close to the transition.  We note
that the
insulating $\delta$-doped GaAs/AlGaAs heterostructures that show very
strong response to in-plane magnetic field\cite{shlimak} were also found to
obey ES hopping with a constant prefactor\cite{khondaker99}.

The magnetoresistance of 3D materials that exhibit ES hopping has been
found to be net negative in some cases \cite{CdSe} and positive in others
\cite{absence}.  A large, negative magnetoresistance that depends strongly on
temperature has been attributed \cite{NSS,Imry} to the effect of a magnetic
field on the quantum interference between forward-scattering hopping paths.
However, this process, as well as all others of orbital origin, is not
relevant to the case under consideration, where a magnetic field is applied
parallel to the two-dimensional plane of the electrons and couples only to
the spins.

Korube and Kamimura \cite{KK} have proposed that alignment of the electrons'
spins can give rise to a positive magnetoresistance by suppressing hops
between singly occupied states via the Pauli exclusion principle.  This
mechanism yields a resistivity which increases with increasing field and
saturates when the spins are fully aligned, consistent with the behavior
shown in Fig.~2 for silicon MOSFET's.  Albeit considerably smaller, a
positive component of the magnetoresistance of In$_2$O$_{3-x}$ films has
been attributed to this mechanism\cite{Zvi}.  We point out, however, that
the theory of Korube and Kamimura assumes that the electron-electron
interaction energy is considerably smaller than the disorder energy, a
condition that is unlikely to be satisfied in these high-mobility,
low-density silicon MOSFET's \cite{energies}.  Si and
Varma~\cite{si99} have suggested that the positive magnetoresistance
associated
with suppression of the triplet channel contribution should vary smoothly
across the transition.  Others have
suggested \cite{klapwijk99,phillips98} that the giant magnetoresistance is
due to the breaking of spin singlets in the insulating phase.  It is important
to note that the similarity of the magnetoresistances in the conducting and
insulating phases indicates that they derive from the same or closely
connected
physics, suggesting a mechanism that is not specific to hopping or to
insulators.

Based on transport studies in exceptionally clean p-GaAs/AlGaAs
heterostructures with ``insulating'' densities ($n_s<n_c$), Yoon {\it et al.}
\cite{yoon99} concluded that the insulating phase is associated with the
formation of a Wigner crystal rather than with single-particle localization.
The possibility that the insulating state at low electron (hole) densities is
due to the formation of a pinned Wigner glass was also suggested in
Refs.~\cite{simmons98,pudalov93}.  A very large (many orders of magnitude)
increase in the resistance of dilute Si MOSFET's in perpendicular magnetic
field
observed in Ref.\cite{diorio90} was also attributed to the formation of a
magnetically-induced Wigner glass.  A strong positive magnetoresistance is
obtained by Chakravarty {\it et al.}~\cite{chakravarty99} in a spin liquid
phase, which freezes in a continuous phase transition to a Wigner glass;
within
this model, the magnetic properties are continuous across the transition
and the magnetoresistance remains positive in the insulating phase.

To summarize, the magnetoresistance of the 2D system of electrons in silicon
MOSFET's varies continuously across the metal-insulator transition,
exhibiting unexpected behavior in the insulating phase.  The
response to a magnetic field applied parallel to the plane of the electrons
is dramatic and entirely similar to that found earlier in the conducting
phase,
indicating that the anomalous behavior associated with the electrons'
spins is a general feature of very dilute, strongly interacting 2D electron
systems.

We are grateful to S.~Bakker and R.~Heemskerk for their contributions in
developing and fabricating the MOSFET's.  We thank V.~Dobrosavljevic and
B.~I.~Shklovskii for useful suggestions.  This work was supported by the US
Department of Energy under Grant No.~DE-FG02-84ER45153.  Partial support was
also provided by NSF Grant No.~DMR~98-03440.

\end{multicols}

\begin{thebibliography}{10}
\bibitem{energies} In silicon MOSFET's the Coulomb repulsion energy,
$U_c\sim e^2(\pi n_s)^{1/2}/\epsilon$, is about 10~meV at
$n_s=10^{11}$~cm$^{-2}$.  Based on the fact that the Landau levels
are well resolved when the magnetic field-induced splitting between them
is on the order of 0.1~meV, the disorder energy $W$ is estimated to be
no more than a
few tenths of meV.
\end{thebibliography}
\end{document}